\documentclass{emulateapj_old}

\def \th{\thinspace}

\def \eg{{{\it e.g.},\ }}
\def \etal{{\it et al.\ }}

\def \cf{{\it cf.\ }}

\def \ie{{{\it i.e.},\ }}

\def \vs{{\it vs.\ }}
\def \Teff{{$T_{\rm {ef\!f}} $}}
\def \teff{{T_{\rm {ef\!f}} }}

\def\ML{{$M$-$L$\ }}
\def\negth{\negthinspace}

\def \aanda{A\&A}

\def \approxgt{\,\raise2pt \hbox{$>$}\kern-8pt\lower2.pt\hbox{$\sim$}\,}
\def \approxlt{\,\raise2pt \hbox{$<$}\kern-8pt\lower2.pt\hbox{$\sim$}\,}

\long\def\jumpover#1{{}}

\slugcomment{to be published in ApJ Letters}

\tablewidth{4cm}

\begin{document}

\title{Detection of Beat Cepheids in M33  and Their
Use as a Probe of the M33 Metallicity Distribution}

\author{J.-P. Beaulieu\altaffilmark{1},
J.~Robert Buchler\altaffilmark{2},
J.-B. Marquette\altaffilmark{1}.
J.D. Hartman\altaffilmark{3},
A. Schwarzenberg-Czerny\altaffilmark{4}
}

\altaffiltext{1}{Institut d'Astrophysique de Paris, 75014
Paris, France}
\altaffiltext{2}{Physics Department, University of Florida,
Gainesville, FL 32611, USA}
\altaffiltext{3}{Harvard-Smithsonian Center for Astrophysics,
Cambridge, MA 02138, USA}
\altaffiltext{5}{Nicolaus Copernicus Astronomical Center,
00-716 Warszawa, Poland}

\begin{abstract} 

Our analysis of the Deep CFHT M33 variability survey database has uncovered 5
Beat Cepheids (BCs) that are pulsating in the fundamental and first overtone
modes.  With {\it only} the help of stellar pulsation theory and of
mass--luminosity ($M$-$L$) relations, derived from evolutionary tracks, we can
accurately determine the metallicities $Z$ of these stars.  The [O/H]
metallicity gradient of $-0.16~\mathrm{dex/kpc}$ that is inferred from the M33
galacto-centric distances of these Cepheids and from their 'pulsation'
metallicities is in excellent agreement with the standard spectroscopic
metallicity gradients that are determined from H II regions, early B supergiant
stars and planetary nebulae.  Beat Cepheids can thus provide an additional,
{\it independent} probe of galactic metallicity distributions.

\end{abstract}




\keywords{
(stars: variables:) Cepheids,
stars: oscillations (including pulsations),
galaxies: individual: M33,
galaxies: abundances,
galaxies: evolution
}

\maketitle

\section{Introduction} \label{sec:intro}

The determination of the abundance distributions provides an important
constraint for evolution models of galaxies.  These abundances which are
traditionally obtained through spectroscopic means, although other approaches
have also been developed, are beset with large uncertainties.  In this Letter
we discuss a totally independent determination of the galactic metallicity ($Z$)
distribution with the help of Beat Cepheids (BCs hereafter) which promises to
have a good accuracy.  Our approach is sensitive to the opacity, and as such it
is sensitive to the iron group elements, but it is relatively insensitive to
the chemical makeup of $Z$, and to $Y$ as well.
 
Cepheids are young, intermediate mass, bright, periodic variable stars.  While
in the core He burning evolutionary phase, they lie in an area of the HR
diagram, the instability strip where their envelopes are vibrationally unstable
through the kappa mechanism \citep{cox68}.  They undergo radial pulsations,
most of them periodic, in a single mode (either the fundamental mode, F, the
first overtone, O1, or the second overtone, O2).  A small fraction of them,
called Double-Mode Cepheids or Beat Cepheids (BCs), are pulsating {\it
simultaneously} in two radial modes.  These pulsations are observed to be
either a mixture of the fundamental and the first overtone modes F/O1, or of
the first and the second overtones O1/O2.  It useful to plot the period ratios
\vs period, a Petersen diagram (PD) named after \citep{petersen73}.  It is well
known that a comparison in a PD of the BCs in our Galaxy, LMC and SMC shows
that the period ratios of F/O1 depend strongly on metallicity~~ (unlike O1/O2
BCs).  This behavior has been taken advantage of to constrain stellar models,
(\eg \cite{mkm92}, \cite{cdp95}, \cite{morganwelch97}) or to test galactic
distances (\eg \cite{kovacs00}, \cite{bbk01}).
 

The M33 late spiral galaxy is known to exhibit a fairly steep metallicity
distribution.  Since \cite{vilchez88} various groups have used HII regions to
estimate the [O/H] distribution in M33.  \cite{garnett97} found an [O/H]
gradient (i.e. abundance versus deprojected galacto-centric radius) of --0.11
dex/kpc in agreement with earlier work.  The first studies with B type
supergiant stars yielded a gradient of $-0.16 \pm 0.06$ \citep{monteverde97},
although a more recent estimatation \citep{urbaneja05} gave $-0.06 \pm 0.02$,
which is also the value obtained through an analysis of red giant branch
photometry by \cite{tiede04}.  From planetary Nebulae \cite{magrini04} derive a
metallicity gradient of --0.14.  A recent study of HII regions by
\citet{crockett06} however revises the [O/H] gradient to $-0.012\pm 0.011$
dex/kpc, much shallower than all the other ones.  As the data of the above
references show, one should keep in mind though that representing the
metallicity distribution as a gradient may be a gross oversimplification.

The M33 variability survey \citep{hartman06} is a wide-field (one square
degree) survey of this galaxy, conducted with the Canada-France-Hawaii
Telescope, in the $g'r'i'$ filters, and the first two years of data are
available with more than 80000 light curves.  This survey is particularly well
suited for variable stars with periods of the order of a few days to a few
weeks.  The excellent relative photometry obtained with the image subtraction
technique \citep{alard00} has given us the opportunity to search for BCs at 840
kpc.  These BCs allow us then to determine the metallicity gradient across the
M33 galaxy.

\section{Beat Cepheids in M33} \label{sec:properties}

\subsection{The Search for Beat Cepheids in M33}

A set of 3023 Cepheid candidates was identified from \citet{hartman06} and
periods were determined simultaneously using the One Way Analysis of Variance,
AoV \citep{schwarzenberg-czerny89}, as implemented for the EROS project
\citep{beaulieu97}.  This phase dispersion minimization method is based on
strong statistical tests and constitutes a powerful way of searching for
periodic signals of arbitrary shapes, and a confidence level is returned for
each trial period.  Each star has a light curve comprising typically 33
measurements in each Sloan band $g',~r',~i'$.  We chose the $r'$ band data in
our analysis because of the highest signal to noise ratio.

\begin{table}
\caption{\small Beat Cepheids in M33}
\vspace{-4mm}
\begin{center}
\begin{tabular}{c c c c c c c}
\hline\hline
ID &\ Number \negth& $\alpha_{2000}$ & $\delta_{2000}$ & $P_0$ & $P_1$ &  $P_{10}$ \\
   &        &                 &                 & (days)& (days)   &                \\
\hline
    \noalign{\smallskip}
A &  121029 & \negth 01:34:59.72 & 30:52:25.2 & 4.70497 & 3.38510 & 0.7195 \\
B &\negth  160520 & \negth 01:32:56.82 & 30:41:33.8 & 3.97755 & 2.86107 & 0.7190 \\
C &\negth  133449 & \negth 01:34:33.43 & 30:51:15.6 & 3.82707 & 2.71407 & 0.7091 \\
D &\negth  234922 & \negth 01:33:54.63 & 30:35:19.8 & 6.17640 & 4.33313 & 0.7015 \\
E &\negth  237367 & \negth 01:34:03.97 & 30:38:08.4 & 6.18792 & 4.33481 & 0.7005 \\
    \noalign{\smallskip}
\hline
\noalign{\flushleft Identifications, coordinates, period of
fundamental mode ($P_0$); of first overtone mode ($P_1$)
and period ratio ($P_{10}$).}  
\end{tabular}
\label{tab:beat}
\end{center}
\vspace{-5mm}
\end{table}

\begin{figure*}
\epsscale{0.95}
      \plotone{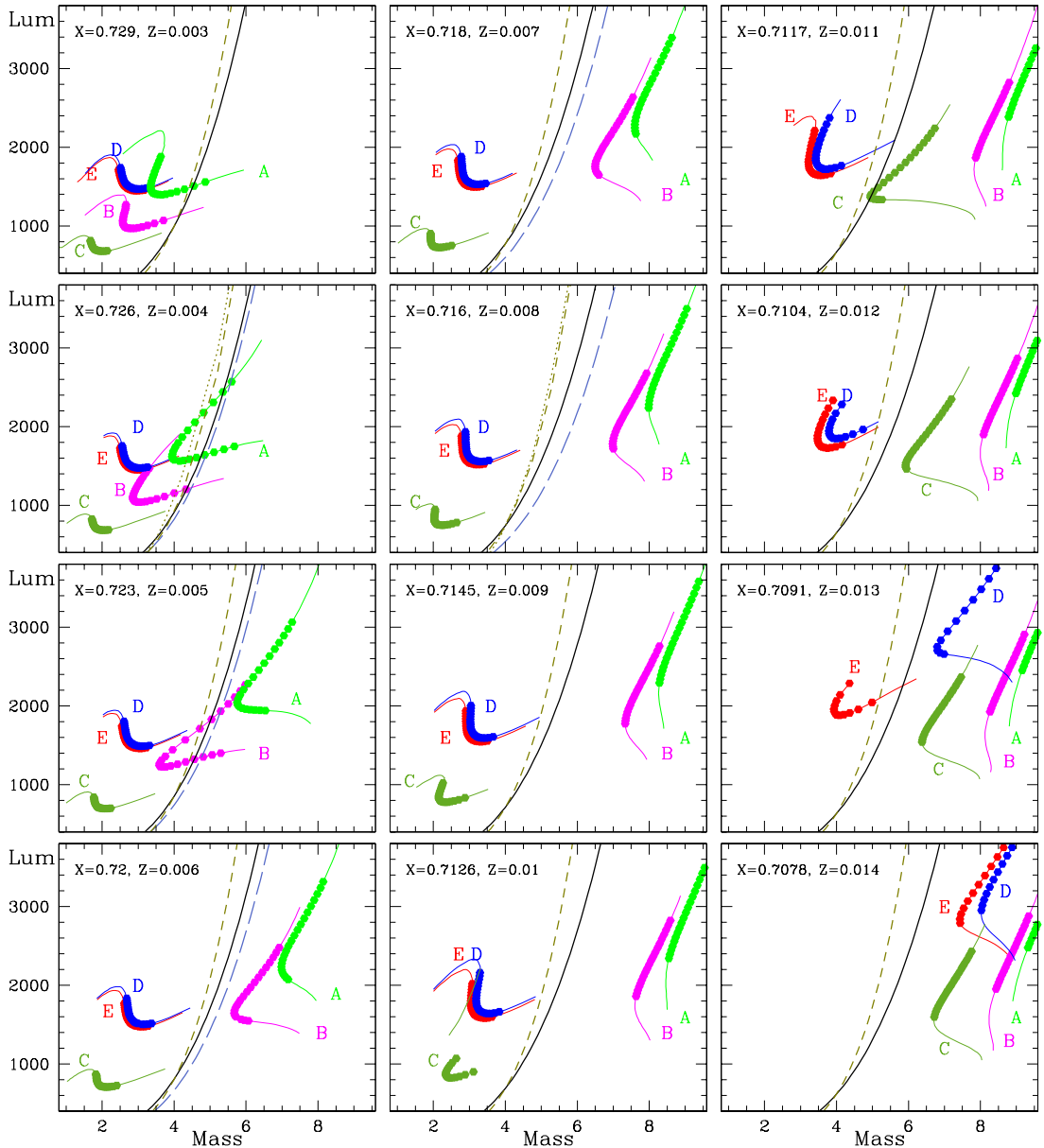}

\caption{Mass-Luminosity plane for 12 different metallicities $Z$ and
corresponding $X$.  Each panel shows the possible solutions (thick dots) for
models with simultaneously unstable F and O1 modes for our 5 BC stars. The
curved lines denote various evolutionary \ML relations (\cf text).{\small ( A
color version of this figure is available from the e-paper)}
}
\label{fig_ml}
\end{figure*}

For each Cepheid candidate, we fitted a Fourier series to the light curve with
the most probable period.  This series was then subtracted and an AoV
periodogram of this prewhitened light curve was computed following the method
described by \citet{beaulieu97}.  For BCs pulsating in the F and O1 modes the
period ratios are known to be $P_{10}=P_1/P_0 \approx 0.7$, whereas for those
pulsating in the O1 and O2 modes it is $P_1/P_2 \approx 0.8$.  We
visually inspected the AoV periodograms of both the prewhitened and of the full
light curves to check for an additional peak near the following periods
$P$/0.7,~$P$$\times$0.7,~$P$/0.8,~$P$$\times$0.8.

In this analysis we uncovered 5 F/O1 type BCs, and their properties are listed
in Table~\ref{tab:beat}.  (We did not detect any O1/O2 BCs.)  The smallness of
this sample reflects the difficulty of searching BCs with only 33 measurements,
and the sample is certainly not complete.  Additional photometric data (at
least 100 measurements) would be needed to significantly increase our detection
rate of M33 BCs.

We note that these 5 BCs span deprojected galacto-centric M33 distances from 1
to 3.5 kpc.  In Table~\ref{tab:model} we also indicate the values of the
metallicities they would have with the standard choice of the [O/H] metallicity
gradient --0.11 dex/kpc, \eg \cite{garnett97} and with the revised one of
--0.012 of \cite{crockett06}, after conversion from [O/H] to [Fe/H]=1.417
[O/H]) as in \citet{maciel03} and to $Z$, using $Z_\odot=0.017$.



\subsection{Pulsational Beat Cepheid Models as Function of Metallicity}

For the purposes of pulsations a chemically homogeneous stellar envelope is
uniquely characterized by three astrophysical parameters in addition to its
composition variables.  We will assume here that the helium content $Y=Y(Z)$ is
a function of $Z$, as one would expect from galactic evolution.  We will find
that our results are relatively insensitive to the exact relation $Y(Z)$.
Furthermore, while the stellar pulsation frequencies are very sensitive to the
abundance of Fe group metals, they are not sensitive that of the remaining
elements which we assume to be a solar mixture.  For computational purposes the
parameters that characterize a Cepheid envelope are most conveniently taken to
be the stellar mass $M$, the luminosity $L$, the effective temperature \Teff\
of the equilibrium model, and the metallicity $Z$.  The linear periods are then
functions of these variables only, \ie $P_k = P_k(L, M, \teff, Z)$.

\begin{table}
\caption{\small Metallicities of M~33 Beat Cepheids}
\vspace{-4mm}
\begin{center}
\begin{tabular}{l l l l l l l l }
\hline\hline
ID \quad &  $P_{10}$ & dist  & [O/H] & \quad $Z$  & [O/H] & \quad  $Z$  &  $Z$      \\
   &               & (kpc) &(G97)\quad  &   (G97)\quad\quad  & (C06)\quad  & (C06)\quad\quad  & from BCs \quad  \\
\hline
    \noalign{\smallskip}
A & 0.7195 & 3.5 & --0.355 & 0.005 & --0.492 & 0.003 & 0.004 \\
B & 0.7190 & 3.1 & --0.311 & 0.006 & --0.487 & 0.003 & 0.005 \\
C & 0.7091 & 2.3 & --0.223 & 0.008 & --0.478 & 0.004 & 0.011 \\
D & 0.7015 & 1   & --0.080 & 0.013 & --0.462 & 0.004 & 0.0125 \\
E & 0.7005 & 1   & --0.080 & 0.013 & --0.462 & 0.004 & 0.0135 \\
    \noalign{\smallskip}
\hline
\noalign{\flushleft Period ratio ($P_{10}$); 
deprojected galacto-centric distance;
[O/H] and $Z$ metallicities from Garnett \etal (1997);  
same for Crockett \etal  (2006);  
$Z$ from the BCs.}
\end{tabular}
\label{tab:model}
\end{center}
\vspace{-5mm}
\end{table}

\begin{figure} 
\epsscale{1.15}
              \plotone{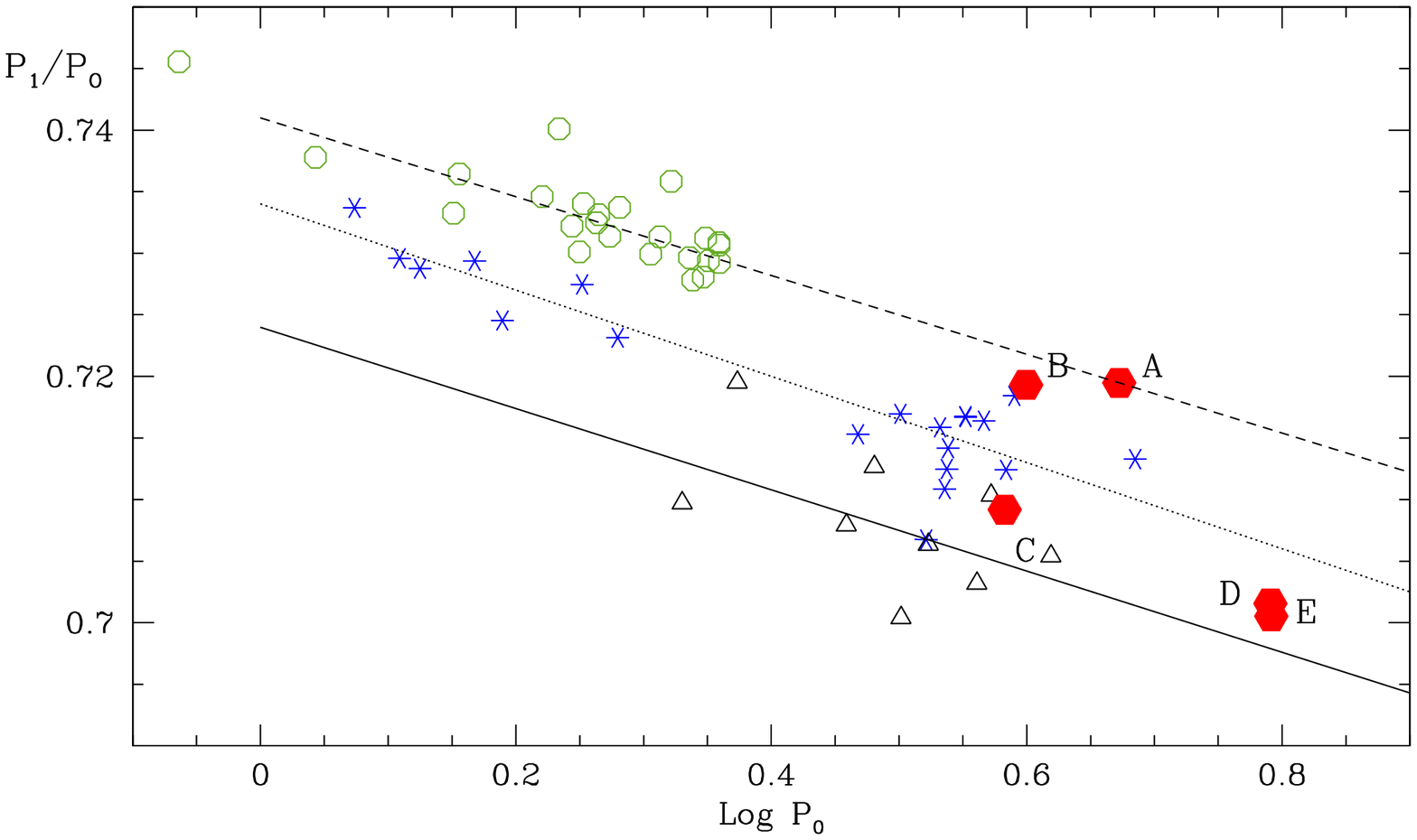}
\caption{Petersen diagram for Beat Cepheids pulsating in
the fundamental and first overtone mode in our Galaxy (open triangles), LMC
(stars), SMC (open circles) and M33 (filled hexagons).  The linear fits are
displayed for our Galaxy (solid), LMC (point) and SMC (dash).}
\label{fig:prat}
\end{figure}

The stellar equilibrium models are constructed and their linear (nonadiabatic)
periods and growth rates computed with the Florida pulsation code.  Turbulent
convection is approximated with a time-dependent mixing length model (Yecko,
Koll\'ath \& Buchler 1998) and the values of the $\alpha$ parameters as defined
in \citet{kollath02} are taken to be \{$\alpha_d$= 2.177, $\alpha_c$= 0.4,
$\alpha_s$= 0.433, $\alpha_n$= 0.12, $\alpha_t$= 0.001, $\alpha_r$= 0.4,
$\alpha_p$= 0, $\alpha_\lambda$= 1.5\}. OPAL \citep{iglesias96} and Alexander
\& Ferguson (1994) opacities are used.

For each BC in our sample we compute a sequence of models with given \Teff\ in
a reasonable range of values, say from 5,300\thinspace K to 6,600\thinspace K.
For each of these \Teff\ we iterate on $L$ and $M$ in order to generate the
observed periods $P_0$ and $P_1$.  The acceptable models are those for which
both the F and O1 modes are linearly unstable.  We note here that the use of
linear periods is adequate because nonlinear periods differ at most in the
fourth decimal figure (\eg \cite{aa98}).

In Fig.~\ref{fig_ml} we display for each of the five BCs (sections of) the loci
of the acceptable stellar models that satisfy the observed periods $P_0$ and
$P_1$ in an \ML diagram.  The large dots denote the acceptable models that are
linearly unstable simultaneously in the F and O1 modes.  Each model along the
loci has a different \Teff.  The adopted composition values, $X$ and $Z$, are
indicated in each subfigure.

Not all values can represent actual Cepheids because the latter move along
evolutionary tracks.  Stellar evolution calculations show that in first
approximation these tracks are horizontal in an HR diagram, \ie $M$ = $M$($L$).
This allows us to narrow down the acceptable BC models.  In Fig.~\ref{fig_ml}
we therefore superpose a sample of evolutionary \ML relations.  The solid
curves represent those of \citet{bono00} as a function of $M$, $Y$ and $Z$.
The short and long dashed curves are derived from \citet{girardi00} and
\citet{alibert99}, respectively, some of them interpolated to the appropriate
composition.  This set of \ML curves should bracket the uncertainties that are
inherent in the evolutionary track calculations.  Note that in the $M$-$L$
diagram, in contrast to the
loci the \ML curves are relatively insensitive to $Z$.

One notes that for each star the width $\Delta M$ (and $\Delta L$) of the
models that satisfy the period constraints turn out to be narrow and comparable
to the uncertainty in the evolutionary \ML relations.  One also notes that as
the metallicity increases, the curves systematically move to the right (to
larger $M$) in a similar fashion for each star.  The loci first shift very
slowly rightward, but once in the broad vicinity of the \ML curves, they
accelerate and cross this region very rapidly, and then slow down again.
This very fortunate feature allows us to put tight limits on the metallicity of
each star, despite the spread in the \ML relations and despite the spread in
\Teff.

From Fig.~\ref{fig_ml} we can therefore read off the metallicities of the 5
stars.  Star C, for example,  crosses the \ML relations
when Z=0.011.  Considering that for Z=0.010 and Z=0.012 star C is comfortably
on either side of the possible \ML relations, we estimate conservatively that
our determination of $Z$ is better than $\pm$ 0.001.  We have verified that a
small change of $Y$ away from the used values has only a minor effect on the
loci.  The metallicities that we infer this way for each star are reported in
Table~\ref{tab:model}.  

Hydrodynamical simulations and the amplitude equation formalism demonstrate
that beat pulsations can occur only in a subregion of the region where the F
and the O1 modes are linearly unstable (e.g. \cite{kollath98, kb01,
kollath02}).  It may therefore be possible to further narrow down the
metallicity of BCs, but this will require an extended survey with nonlinear
numerical hydrodynamical simulations.

Table~\ref{tab:model} also displays the metallicities that are inferred for our
BCs from the 'standard' metallicity gradient, \eg \citet{garnett97}.  The
agreement is found to be excellent although star C has a somewhat high $Z$ in
our study.  On the other hand, our results are incompatible with those derived
from the revised gradient of \citet{crockett06}.

From our $Z$ values and from the galacto-centric distances of the 5 BCS we
derive a metallicity gradient d{\th}Log($Z$)/dR = --0.2 dex/kpc which
translates to an [O/H] gradient of --0.16 dex/kpc that is in good agreement
with the standard determinations, and rules out the shallow revised one.

\subsection{Empirical Determination of the Metallicity}

Let us reexamine our derived metallicities in light of Fig.~\ref{fig:prat} in
which we have plotted a PD for all the known F/O1 Beat Cepheids in the Galaxy,
in the LMC \citep{alcock95,soszynski00} and in the SMC
\citep{beaulieu97,udalski99}.  We also show the best linear fits for all 3
galaxies (although there is no theoretical reason for the constant $Z$ curves
to be straight lines).  Our 5 BCs from M33 are plotted as solid hexagons.  The
2 stars A and B, with the highest period ratios, lie approximately in the
continuation of the SMC Cepheids, and indeed their derived metallicities
of $Z\sim$0.004 and $Z\sim$0.005 agree with that of the SMC ($Z$=0.004).  The
star C with a derived $Z=0.011$, falls between the LMC ($Z$=0.008) and the
Galactic BCs ($Z$=0.020).  Finally, the remaining two stars, D and E, have the
lowest period ratios, 0.7015 and 0.7005, and with derived $Z=0.0125$ and
$Z=0.0135$, lie between the LMC and Galactic BCs.

It is important to stress that Fig.~\ref{fig:prat} uses {\sl only observed
quantities}, namely periods and is therefore independent of any theoretical
modeling.  With a little act of faith one could have 'predicted' the $Z$ of the
5 M33 BCs, although for 3 of these 5 stars a huge extrapolation would have been
required.  Our calculations suggest that such an extrapolation, whether linear
of curved, may be warranted.  However, it will be necessary to make a detailed
survey of linear and nonlinear BC models to ascertain with what precision the
mere location of a BC in the PD is able to provide an empirical estimate of its
metallicity.  It has been suggested that stellar rotation may falsify
the use of the PD as a metallicity indicator as for delta Scuti stars
\citep{sgg06}.  We have checked that for Cepheid models, and for their lower
rotation rates, the rotational effect on the PD is negligible.

Fig.~\ref{fig:prat} shows that 4 of the 5 M33 BCs have extremely large periods
when compared to the known MC and Galactic BCs.  Clearly, the faintness of M33
precludes the detection of low luminosity (and thus short period) Cepheids.
But why are the MCs and the Galaxy devoid of such large period BCs?  It is not
clear at this time whether this is a selection effect or an evolutionary effect (e.g. the horizontal
tracks not extending far enough blueward) or a pulsational effect (the
occurrence of beat behavior is fragile in the sense that it can be easily
affected by small changes in astrophysical parameters such as the chemical
makeup), or whether some other effect is responsible.

\vskip 0.5 cm

To finish it is worth reflecting on why this procedure of determining the
metallicity of BCs merely on the basis of the two observed periods works so
well.  There are two reasons:~~First, Cepheids evolve along tracks, \ie
$L=L(M(t),\teff(t), Z(t))$; to the extent that these are approximately
horizontal we can assume a \ML relation in the form $L=L(M,Z)$;~~second, the
subclass of the BCs lies in a narrow substrip of the Cepheid instability strip
(\eg Koll\'ath \etal 1998, 2001, 2002); thus, for a given $M$ and metallicity
$Z$, the effective temperature is restricted to lie in a narrow range, $\teff
\approx \teff (M,L,Z)$.  These two constraining relations, plus those of the
two measured periods therefore provide the 4 relations that are necessary to
uniquely specify the Cepheid model ($L$, $M$, \Teff) and its composition $Z$.

\section{Conclusions}

We have searched for Beat Cepheids in the database of the Deep CFHT M33
Variability Survey, and we have uncovered 5 of them that are pulsating in the
fundamental and first overtone modes.  With the use of stellar pulsational
modeling and evolutionary \ML relations we have derived the metallicities of
these BCs.  From the galacto-centric distance $R$ of the BCs we have inferred a
'pulsational' metallicity gradient d{\th}Log $Z$/dr = --0.2 dex/kpc that
translates to an [O/H] metallicity gradient of --0.16 dex/kpc, in good
agreement with the other determinations in the literature, but it rules out the
revised value of \cite{crockett06}.  We stress that this 'pulsational' method
of probing the galactic metallicity distribution is independent of
spectroscopic or other techniques that have previously been used.  \\


This work is based on observations with Mega Prime/MegaCam, a joint project of
CFHT and CEA/DAPNIA, at the Canada-France-Hawaii Telescope (CFHT) which is
operated by the NRC of Canada, the Institute National des Sciences de l'Univers
of the CNRS of France, and the University of Hawaii.  We express our gratitude
to the CFHT staff at Mauna Kea for conducting the service mode observations on
which this study is based.  This work has been supported by NSF (AST03-07281)
at UF.  One of us (JRB) has profited from the 2006 Aspen Center for Physics
workshops on galactic evolution.  He also gratefully acknowledges the
hospitality of the IAP.



\end{document}